\documentclass[aps,prl,twocolumn,superscriptaddress,showpacs]{revtex4}
\usepackage{epsfig}
\begin{document}
\title{Statistical Properties of The First Excited State of an
Interacting Many Particle Disordered System}
\author{Richard Berkovits}
\affiliation{The Minerva Center, Department of Physics,
    Bar-Ilan University, Ramat-Gan 52900, Israel}
\author{Yuval Gefen} 
\affiliation{Department of Condensed Matter Physics,
    The Weizmann Institute of Science, Rehovot 76100, Israel}
\affiliation{The Minerva Center, Department of Physics,
    Bar-Ilan University, Ramat-Gan 52900, Israel}
\author{Igor V. Lerner}
\affiliation{School of Physics and Astronomy, University of Birmingham,
Birmingham~B15~2TT, United Kingdom}
\author{Boris L. Altshuler}
\affiliation{NEC Research Institute, 4 Independence Way, Princeton, New Jersey 08540}
\affiliation{Department of Physics, Princeton University, Princeton, New Jersey
08545 }
\date{Mar. 3, 2003, version 3.1}
\begin{abstract}

The distribution $P_1$  of the first many-body  excitation energy 
of a weakly and moderately  interacting electron gas in a finite
conductor (in the diffusive regime)  is calculated. As the interaction
is increased, $P_1$  crosses over from Wigner-Dyson to Poisson. We characterize this
transition  through the inverse participation ratio in Hilbert space, and 
examine its manifestation in a projected $2$-dimensional  space.

\end{abstract}
\pacs{PACS numbers: 71.30.+h, 64.60.Ak, 73.20.Fz}
\maketitle

One of the most important physical quantities used to characterize
finite conductors is  energy level statistics. Not only does it provide an
efficient mathematical framework  to study such systems, but it is also
correlated with   other physical observables  (conductance, spectral 
correlations,  absorption spectrum, orbital magnetization, heat capacity 
when the number of  electrons is fixed,  etc.). 
This, in principle,  allows for the ``measurement 
of level statistics''. In weakly disordered conductors 
(``diffusive disorder'', as well as in chaotic cavities),
it is known that the single electron energy levels obey the Wigner-Dyson
 (WD) statistics.

One can further  investigate the 
many-body spectra of such systems. 
Let us first take the interaction strength to vanish. In this case
the  distribution  of the normalized  energy gap between the 
ground-state ($|0\rangle$) 
and the first-excited  
($|1\rangle$), $s\equiv(E_1-E_0)/\langle E_1-E_0 \rangle$,  follows the WD
distribution  (a single particle-hole pair  at the Fermi energy is involved).
Hereafter 
$\langle \ldots \rangle$ denotes  an  average taken 
 over disorder realizations.
Once we consider the gap between the $m^{th}$  and the $m+1^{st}$ 
 many-body states with $m>>1$, we expect the  
statistics to follow the Poisson
distribution. This is due to the fact that, typically, the 
  $m^{th}$  and the $m+1^{st}$  many-body states correspond to mutually
very different particle and hole  configurations. Since they   are 
not connected with  each other through a single-particle operator, 
level repulsion is greatly reduced.  Once we turn on electron-electron 
interaction, we  expect  the gap between 
the $m^{th}$  and the $m+1^{st}$ 
 many-body states ($m>>1$) of such a system  to  satisfy the WD 
distribution. Indeed,  the motivation  behind the introduction of random matrix
theories was to account for high-lying spectra of interacting systems. 
 Table 1 
summarizes the existing  knowledge  regarding the statistics of the excitation
spectrum of  a non-integrable electron gas. For the sake of definiteness
we shall consider hereafter spinless electrons moving under the influence of weak
disorder which renders the motion of a single electron diffusive.

The item  in Table 1 which is marked by a question mark represents
the distribution of the first excitation energy of a 
(weakly or moderately weakly)
interacting electron gas, $P_1(s)$. This quantity  is indeed the main 
object of our
investigation. 
Earlier studies \cite{pichard} of $P_1$ have 
addressed different parameter regimes,
or  were at times inconclusive.

\begin{table} [t]
{\protect 
\begin{tabular}{|c|c|c|}
\hline
     & non-interacting & interacting \\ \hline
first excitation  & Wigner-Dyson & ? \\
higher excitations  & Poisson & Wigner-Dyson \\ \hline
\end{tabular}}
\caption{\small The  many-body spacing statistics of diffusively  systems: 
the item marked by ? is found to cross-over  ( with increasing interaction strength) from WD to Poisson}
\end{table}

\begin{figure}\centering
\epsfxsize7.5cm\epsfbox{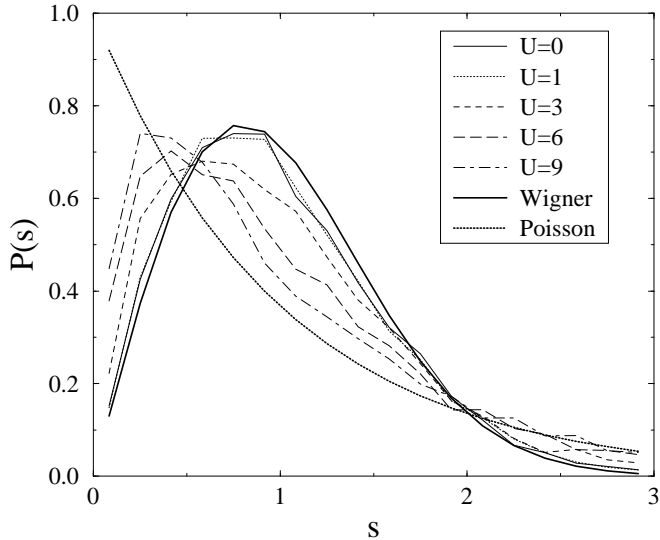}
\caption{The first excitation distribution $P_1(s)$ for different values of
the interaction strength $U$. The Wigner-Dyson  and Poisson distributions are
plotted for comparison.}
\label{fig1}
\end{figure}

\begin{figure}\centering
\epsfxsize7.5cm\epsfbox{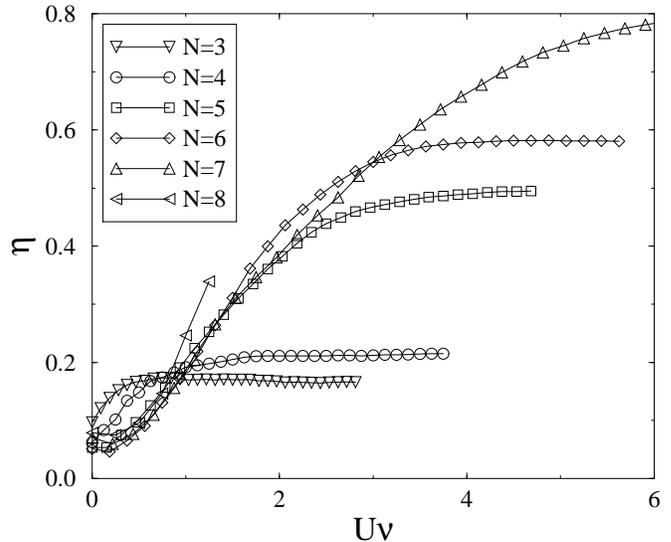}
\caption{$\eta_1$ (cf.  Eq.\ref{eq:eta}) as function of the scaling variable
$U \nu$ for various  values of the filling factor $\nu$. 
Other than  at low density ($\nu \leq 1/4$) and at half-filling the data
  seem to collapse on the same curve with  $\nu$-specific 
saturation values.}
\label{fig3}
\end{figure}

\begin{figure}\centering
\epsfxsize7.5cm\epsfbox{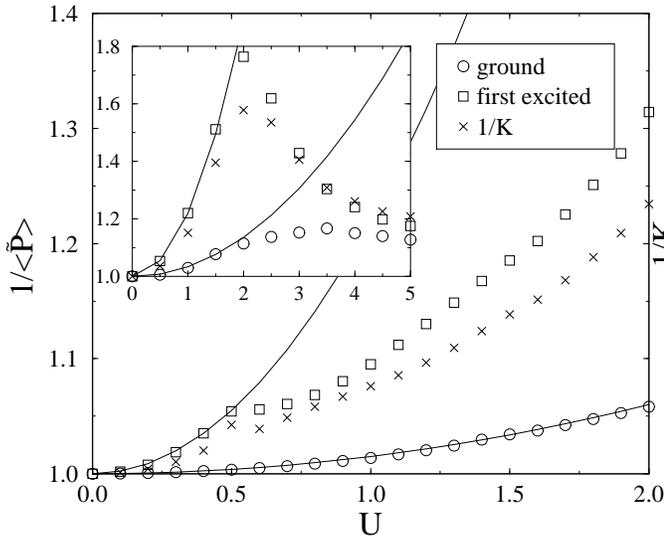}
\caption{The inverse averaged participation ratio  $1/\langle {\tilde  P} 
\rangle$ and $1/K \equiv  1/\langle 1| c_{N+1}^{\dag} c_{N}|0 \rangle$,
as a  function of the interaction strength $U$ for the ground state and the 
first excited state for the $N=4$ case. The curves correspond to the best
fit to $1+aU^2$, with $a=0.015$ for the ground state and
$a=0.22$ for the first excited one. Inset:
The same as the  above for $N=7$ and a wider range of interaction
of interaction strength.  Here
$a=0.034$ for the ground state and
$a=0.22$ for the first excited one.}
\label{fig4}
\end{figure}

The main results   of our analysis are:

\begin{enumerate}
\item We find numerically that as the electron-electron interaction
strength is increased, the statistics of the first excitation energy crosses
over from  WD to Poisson (cf. Fig. \ref{fig1}). 
We characterize this transition 
quantitatively (cf. Fig.\ref{fig3} and  Eq.\ref{eq:eta}).
\item We relate the WD-to-Poisson transition to  the statistics of 
the diagonal and off-diagonal elements of an effective $2 \times 2$ matrix,
depicting the  mixing of the many body-sates $|0\rangle$ 
and  $|1\rangle$ (Fig.\ref{fig8} 
and Eqs. \ref{eq:matrix_elements},\ref{2lev},\ref{2levex}). 
\item We show how the  WD-to-Poisson crossover is manifest in the distribution
of the participation ratio, i.e., the structure of the 
many-body wave function in Hilbert space.
\item Finally, and very importantly, we show that as the interaction is
increased the first excited state, which  originally (at zero interaction) 
corresponds to a single  particle-hole excitation, 
involves an increasing number
of particle-holes. But this is up to a certain  value  of the interaction
($r_s \approx 1$). Upon  
further increase of the interaction strength this trend reverses. This picture is
obtained  by computing
 (numerically) the inverse
participation ratio of the ground and first-excited states (in Hilbert space)
as function of the interaction (cf. Fig. \ref{fig4})
\end{enumerate}

{\it The model.}
We consider the following
tight-binding Hamiltonian, describing spinless interacting 
electrons moving in the presence of a random potential:
\begin{eqnarray}
{\hat H} = \sum_{k,j} \epsilon_{k,j} n_{k,j}
- V \sum_{k,j} [a_{k,j+1}^{\dag} a_{k,j} +
a_{k+1,j}^{\dag} a_{k,j} + h.c. ] \nonumber \\
+ U \sum_{k,j} n_{k,j} n_{k \pm 1,j \pm 1} \,\, .
\label{hamil}
\end{eqnarray}
Here $\{k,j\}$ denotes a lattice site, 
the number operator is
$n_{k,j}$
and
$\epsilon_{k,j}$ is the site energy, selected randomly and uniformly 
over the interval  $[-W/2,W/2]$.
We study two-dimensional arrays with periodic boundary conditions.
Only nearest-neighbor interactions $U$ are considered \cite{fn0,fn1}.

We present  numerical results  pertaining 
to  $N=4 - 8$ electrons residing on
$4 \times 4$ lattices with $M=16$ sites ($\nu \equiv N/M$). Data for larger systems appeared to 
be compatible with our results.
 The disorder strength $W=8V$
was chosen so that the
single-particle localization length is larger than the system
size, with  spectral correlations 
following  the 
Gaussian orthogonal 
ensemble (GOE) on small energy scales (i.e., for energy differences
$\Delta \epsilon < \hbar /t_{flight}$ ; $t_{flight}$ is the 
time-of-flight of the electron  throughout the system).  The 
 motion of the electron is diffusive or marginally diffusive,
$l_{el} \leq L$, with $L$ being the quantum dot's linear 
size and $l_{el}$ -- the elastic
mean-free-path.)
Varying the 
electron density in the experiment  amounts to varying the dimensionless 
parameter $r_s$, 
\cite{fn3}.
We have carried  out exact diagonalization of the many-particle
Hamiltonian using the Lanczos method, and obtained  the eigenvalues
$E_{\alpha}$ and eigenvectors $|\alpha\rangle$,
 $\alpha=0,1$.

To establish a useful  frame of reference for discussing the results
of our analysis, it is convenient to study the 
self-consistent Hartree-Fock (SCHF) Hamiltonian corresponding to this
system. This Hamiltonian reads
\begin{eqnarray}
\label{eq:shf}
\hat H_{HF}=\sum_{k,j} \epsilon_{k,j} n_{k,j}
- V \sum_{k,j} [a_{k,j+1}^{\dag} a_{k,j} +
a_{k+1,j}^{\dag} a_{k,j} + h.c. ] \nonumber \\
\sum_{k,j} n_{k,j} U \langle n_{k \pm 1,j \pm 1} \rangle_0
-\sum_{k,j} a_{k,j}^{\dag} a_{k \pm 1,j \pm 1} U
\langle a_{k \pm 1,j \pm 1}^{\dag} a_{k,j} \rangle_0,
\end{eqnarray}
Here $\langle\ldots\rangle_0$ denotes  a quantum average taken over the ground
state, the latter being  calculated self-consistently. The SCHF single-electron
eigenvectors $\{|\psi_n\rangle\}$ and eigenvalues $\{\varepsilon_i\}$ are
obtained through a self-consistent diagonalization of the
Hartree-Fock Hamiltonian.

The exact many-body states
may be expressed in terms of
Fock determinants :
\begin{eqnarray}
\label{fock}
|\alpha\rangle = \sum_{i_1,i_2,\ldots,i_N}
C^{(\alpha)}_{i_1,i_2,\ldots,i_N} c_{i_N}^{\dag}\ldots c_{i_2}^{\dag} c_{i_1}^{\dag}
|{\rm vac}\rangle,
\end{eqnarray}
where the sum is over all  possible permutations of $N$ states out of
$M$ with
$ M \geq i_N>i_{N-1}>\ldots>i_2>i_1 \geq 1$,  $c_n^{\dag}$ is the creation
operator of the $n^{th}$  single-particle  HF state and $|{\rm vac}\rangle$ 
is the vacuum
state. For the non-interacting case, as well as for the case
for which the SCHF provides an exact solution, the ground state
(the first excited state) 
corresponds to  $C_{1,2,\ldots,N}=1$ ($C_{1,2,\ldots,N-1,N+1}=1$) 
while all other coefficients are
equal to zero. This is compatible with a Koopmans-like picture. 
In this case the participation ratio (see below) is equal to $1$.

An important quantity
which is particularly suitable to describe scenarios  intermediate between
the WD ($\eta = 0$) and the Poisson ($\eta = 1$)  limits is \cite{fn2}

\begin{eqnarray}
\label{eq:eta}
\eta_1 = {{\int_0^{0.4729} P_1(s) ds - \int_0^{0.4729} P_{wd}(s) ds}
\over{{\int_0^{0.4729} P_p(s) ds - \int_0^{0.4729} P_{wd}(s) ds}}}\,\, .
\end{eqnarray}

{\it WD to Poisson crossover.}
The distribution of $s$ as function of the
interaction strength for $1000$ different realizations of  disorder 
is plotted in Fig. \ref{fig1}.  It is seen that
 $P_1(s)$ crosses over from a Poisson-like 
 to a WD-like distribution
 as the interaction strength is increased, with the canonical
signatures of such a transition in finite systems.
The functional $\eta$ 
is depicted in  Fig. \ref{fig3}.
Each  curve represents averaging
 over $10000$ different disorder  realizations.  
In all cases $\eta$ increases (the distribution becomes more ``Poisson-like'')
as the interaction strength $U$ is increased.
This  increase in $\eta$ becomes more pronounced  the  higher 
the filling factor, 
and tends to saturate
at higher interaction strengths. 
At   filling factors $\nu >  1/4$ (and $\nu \ne 1/2$)
our  data suggest that $\eta$ 
 collapses  onto a single scaling function 
up to $\nu$-specific values where saturation takes place.

{\it Comparison with a SCHF scheme.}
It is known \cite{AGKL} that in zero-dimensional systems the low-lying 
many-body states are made of combinations of a finite number of Slater
determinants (each made up of single-particle SCHF states). This motivates
the use of such  a basis to describe the exact $|0\rangle$ and $|1\rangle$
states.
We invoke the participation ratio in the Hilbert space spanned
by all possible Slater determinants:
\begin{eqnarray}
\label{eq:eta1}
{\tilde P} =  \sum_{i_1,i_2,\ldots,i_N}
\vert C^{(\alpha)}_{i_1,i_2,\ldots,i_N}  \vert^4.
\end{eqnarray}
Koopmans' variant of the HF approximation asserts that the effective 
single-particle states remain unmodified under a change of the 
occupancy of these states.
As long as the HF-Koopmans-like picture holds  \cite{fn3}
  ${\tilde P} =  1$, 
whereas  if this 
picture breaks down completely we approach  
 ${\tilde P}\sim \left(^{M}_{N}\right)^{-2}$, signaling delocalization
in Hilbert space\cite{AGKL}.

Our results are depicted in Fig. 
\ref{fig4}. There we  plot  the disorder-averaged inverse 
participation ratio as function
of the interaction for the ground-state and the first
excited state, with  $N=4$ and $N=7$. 
It can be seen that 
$1/\langle {\tilde P} \rangle \propto U^{2}$ for 
weak interactions for both states. 

\begin{figure}\centering
\epsfxsize7.5cm\epsfbox{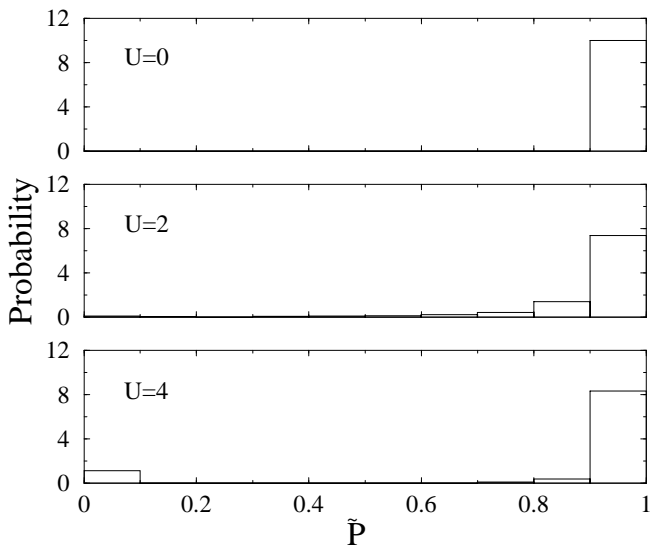}
\caption{The distribution function  of the participation ratio, $\tilde P$, 
for the first excited state.}
\label{fig5}
\end{figure}

\begin{figure}\centering
\epsfxsize7.5cm\epsfbox{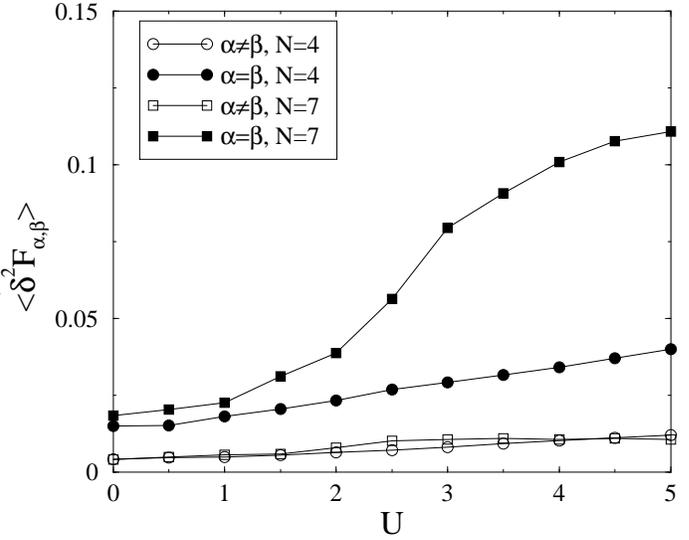}
\caption{The fluctuations in the matrix element,
$F_{\alpha,\beta} = \langle \alpha | \hat F |\beta \rangle$, defined as
$\langle \delta^2 F_{\alpha,\beta} \rangle = \langle F_{\alpha,\beta}^2 \rangle
- \langle F_{\alpha,\beta} \rangle^2$, for the diagonal ($\alpha$=$\beta$)
and non-diagonal ($\alpha\ne\beta$) elements as function of the interaction
strength $U$. For $N=4,7$ $r_s \approx 0.56U, 0.42U$   respectively.}
\label{fig8}
\end{figure}

We note that there is some nontrivial
physics taking place:  the cusp of the curve
 $1/\langle {\tilde P} \rangle$  vs. $U$ for the first excited state, as
it departs from the $\propto  U^{2}$ behavior (at $U=0.5$ for the $N=4$ 
 case).  This cusp becomes more pronounced as $N$ becomes larger; the $N=7$  
case is a manifestation of the  system becoming more
localized ( in Hilbert space)  with  interaction strength exceeding $U = 2$.
\cite{foot}
Studying the distribution of ${\tilde P}$ (cf.  Fig \ref{fig5}) reveals that 
as the interaction strength is increased, the original sharp peak
 at $1$ is initially smeared, but the function then evolves into a doubly
peaked distribution.

In this context it is also useful to consider the  quantity
$K \equiv \langle 1| c_{N+1}^{\dag} c_{N}|0 \rangle$. We note that in the Koopmans
limit $K=1$, implying that both the ground and the first-excited states are given
respectively by a single Slater determinant (they  differ from each other 
by a particle-hole excitation).
In that limit, as discussed above, the inverse  
participation ratio  $1/\langle {\tilde P} \rangle=1$. As we turn on the interaction
$1/\langle {\tilde P} \rangle$ increases concomitantly with the decrease in $K$.
The behavior of the average $K$ is depicted in Fig \ref{fig4}. 

{\it A two-level model}.
To further characterize the phenomena  discussed here, we resort to 
the philosophy behind the Wigner surmise, and try to study the manifestation
of the crossover in a truncated $2 \times 2$ Hilbert space. The 
latter is spanned by 
 $|0\rangle$ and  $|1\rangle$ (in a certain realization).

If the Koopmans-like scenario holds \cite{alhassid01,piechon01,alhassid99}
$K = 1$ while
$\langle 1| c_{N}^{\dag} c_{N}|0 \rangle = 0$. Under such a scenario 
the Wigner distribution should hold: consider an arbitrary  perturbation
$\hat F = \sum_{ij} \alpha_{i,j} c_i^{\dag}c_j$, 
\begin{eqnarray}
\label{eq:matrix_elements}
\langle 0| \hat F |0 \rangle &=& \sum_{i<N+1} \alpha_{ii} \nonumber \\
\langle 1| \hat F |1 \rangle &=& \sum_{i<N} \alpha_{ii} + \alpha_{N+1,N+1}
\nonumber \\
\langle 1| \hat F |0 \rangle &=& \alpha_{N,N+1},
\end{eqnarray}
corresponds (up to a constant matrix  $\hat {\cal{I}} \times \sum_{i<N} 
\alpha_{ii}$ 
subtracted, where $\cal{I}$ is the unit matrix) 
the $2 \times 2$ matrix:
\begin{equation}
\hat F_{HF-Koopmans}=
\left(\matrix{ \alpha_{N+1,N+1} & 
 \alpha_{N,N+1} \cr
\alpha_{N+1,N} & 
\alpha_{N,N} \cr}
\right).
\label{2lev}
\end{equation}
The distribution of the HF states  over  disorder
realizations obeys RMT. Likewise, the distribution of  $\{ \alpha_{ij} \}$
will follow RMT (Gaussian), and will lead to  
  the energy spacing $E_1-E_0$ obeying  the WD distribution.

Going now beyond the Koopmans limit, and employing the basis 
of the exact ground  and first-excited states,  the  2-level
matrix is
\begin{equation}
\hat F=
\left(\matrix{\langle 1| \hat F |1 \rangle & 
\langle 1| \hat F |0 \rangle \cr
\langle 1| \hat F |0 \rangle &
\langle 0| \hat F |0 \rangle\cr}
\right).
\label{2levex}
\end{equation}

The WD $\rightarrow$ Poisson crossover is now manifest through  $\hat F$.
We first subtract a constant, proportional to the unit matrix, to render
the resulting matrix $ \delta \hat F$ statistically  traceless.
We are left to compare the {\it fluctuations} of
the diagonal matrix elements, $\langle \delta^2 F_{\alpha,\beta} \rangle = 
\langle F_{\alpha,\beta}^2 \rangle
- \langle F_{\alpha,\beta} \rangle^2$ with $\alpha=\beta$, with those 
of the off-diagonal entries,   
$\langle \delta^2 F_{\alpha,\beta} \rangle$ with
$\alpha\ne\beta$. (We note that the 
ensemble average of the off-diagonal entries vanishes). 
This is shown in Fig. \ref{fig8} for two
values of electron number, $N$. It is clearly seen, especially for the  larger 
$N$, that as the interaction strength  is increased (beyond $U \approx 2$ which
is equivalent to $r_s=U/V\sqrt{4 \pi \nu}$), 
the diagonal entries win over the 
off-diagonal, marking  the suppression of level-repulsion and the cross-over
to Poisson statistics. This cross-over appears to be quite sharp for our larger
system. 

We acknowledge very useful discussions with and comments by B. Spivak,
and with N. Taniguchi  on possible diagrammatic analysis of this problem.
We acknowledge support from the U.S.-Israel Binational Science Foundation, The
Israel Academy of  Science,  the German-Israel Foundation
and the Minerva foundation.


\begin{thebibliography} {50}

\bibitem{pichard} See e.g. G. Katomeris, F. Selva and J.-L. Pichard, cond-mat/0206404;
G. Benenti, X. Waintal, J.-L. Pichard and D. L. Shepelyansky, Eur. Phy. J. B
{\bf 17}, 515 (2000).


\bibitem{fn0} This choice of
a short range interaction 
is a reasonable way to model a restricted-geometry  
two-dimensional electron gas 
(quantum dot) with metallic gates in the vicinity which give rise 
to screening of the bare Coulomb interaction.
The effects we discuss here  are found for long-range Coulomb interactions
too. In that case, though, one needs to consider higher values of 
the interaction for the effects to be appreciable.

\bibitem{fn1} In principle we need to vary 
the bare potential with U to
keep the ``dressed'' parameter $l_{el}$ unchanged. But  over the range $U$ is
varied here we have observed that $l_{el}$ remains unchanged, the 
evidence for this being that the conductance is hardly modified.

\bibitem{fn3} This is in agreement with :
 P. N. Walker, Y. Gefen and G. Montambaux 
Phys. Rev. B, {\bf 60}, (1999); P. N. Walker, Y. Gefen and G. Montambaux 
Phys. Rev. Lett. {\bf 82}, (1999); A. Cohen, K. Richter, and R. Berkovits, 
Phys. Rev. B , {\bf 60}, 2536 (1999).


\bibitem{fn2} Note also that the
upper limit in the integrals in Eq. \ref{eq:eta} is 0.4729
the value of the lower
intersection point of the WD and the Poisson distributions.
Here we focus on the low energy limit of the distribution function which
is robust and universal. 

\bibitem{AGKL} B. L. Altshuler, Y. Gefen, A. Kamenev and L. S. Levitov,
\prl {\bf 78}, 2803 (1997).

\bibitem{foot} The cusp appears to be weakly geometry dependent.
It is most pronounced for square samples above quarter filling.

\bibitem{alhassid01} Y. Alhassid and Y. Gefen cond-mat/0101461.

\bibitem{piechon01} F. Pi\'echon and G. Montambaux cond-mat/0111433.

\bibitem{alhassid99} Y. Alhassid and S. Malhotra 
Phys. Rev. B , {\bf 60}, R16315 (1999).

\end{thebibliography}
\end{document}